\documentclass[preprint,showpacs,floatfix] {revtex4} 
\usepackage{graphicx}
\usepackage{subfigure}
\usepackage{longtable}
\newcommand{\rvec}{\mathrm{\mathbf{r}}}
\newcommand{\rpvec}{\mathrm{\mathbf{r'}}}
\begin{document}
\title{Density functional studies on the hollow resonances in Li-isoelectronic 
sequence (Z=4--10)}
\author{Amlan K. Roy}
\affiliation{Department of Chemistry, University of New Brunswick, Fredericton,
NB, E3B 6E2, Canada}
\email{akroy@chem.ucla.edu}
\begin{abstract}
In this sequel to our work on triply excited hollow resonances in three-electron atomic 
systems, a density functional theory (DFT)-based formalism is employed to investigate
similar resonances in Li-isoelectronic series (Z=4--10). A combination of the 
work-function-based local nonvariational exchange potential and the popular gradient 
plus Laplacian included Lee-Yang-Parr correlation energy functional is used. The generalized 
pseudospectral method provides nonuniform and optimal spatial discretization of the 
radial Kohn-Sham equation to obtain self-consistent set of orbitals and densities in
an efficient and accurate manner. First, all the 8 n=2 intrashell states of B$^{2+}$, 
N$^{4+}$ and F$^{6+}$ are presented, which are relatively less studied in the literature 
compared to the remaining 4 members. Then calculations are performed for the 8 
$2l2l'$n$l''$ (3$\leq$n$\leq$6) hollow resonance series; {\em viz.,} 2s$^2$ns $^2$S$^e$, 
2s$^2$np $^2$P$^o$, 2s$^2$nd $^2$D$^e$, 2s2pns $^4$P$^o$, 2s2pnp $^4$D$^e$, 2p$^2$ns 
$^4$P$^e$, 2p$^2$np $^4$D$^o$ and 2p$^2$ns $^2$D$^e$, of all the 7 positive ions. Next, as 
an illustration, higher resonance positions of the 2s$^2$ns $^2$S$^e$ series are calculated 
for all the ions with a maximum of n=25. The excitation energies calculated from 
this single-determinantal approach are in excellent agreement with the available 
literature data (for the n=2 intrashell states the deviation is within 0.125\% and 
excepting only one case, the same for the resonance series is well below 0.50\%). With an 
increase in Z, the deviations tend to decrease. Radial densities are also presented for 
some of the selected states. The only result available in the literature for the lower 
resonances (corresponding to a maximum of n=17) have been reported very recently. The n$>$16 
($>17$ for F$^{6+}$) resonances are examined here for the first time. This gives a 
promising viable and general DFT scheme for the accurate calculation of these and other 
hollow resonances in many-electron atoms.
\end{abstract}
\pacs{31.15.-p, 31.50.+w, 32.30.-r, 32.80.Dz}
\maketitle

\section{Introduction}
With all three electrons residing in n$\geq$2 shells leaving the K shell empty, the 
three-electron atomic system constitutes an interesting multi-excited atomic problem. 
These are ideal for examining the delicate interelectronic correlation of three electrons 
under the influence of a nucleus, the well-known four-body Coulombic problem. A large 
majority of these triply excited hollow states are autoionizing and have found important 
practical applications in the field of high-temperature plasma diagnostics. Ever since 
their first observation in an electron-helium scattering experiment by Kuyatt {\em et al.} 
[1] and the same for Li in a beam-foil experiment by Bruch {\em et al.} [2,3], many 
subsequent attempts were made to study the spectra of these challenging systems over 
the past three decades. However the lowest 2s$^2$2p $^2$P$^o$ resonance of Li in a 
photoabsorption experiment was reported by Kiernan {\em et al.} [4] only in 1994 with 
the aid of a dual-laser plasma technique. In recent years, continuing impressive 
developments in the synchrotron radiation technology has spawned a great upsurge of 
interest to measure the position, width and lifetime of these hollow atomic and ionic 
resonances with greater accuracy and precision. Some of these include for example, but not 
limited to, the measurement of (a) the various even- and odd-parity hollow resonances using 
photolectron, photoion spectroscopy [5-9], (b) the photoionization cross section [10], 
(c) the Rydberg series [11] as well as (d) both K and L shell vacancy states, the 
so-called doubly hollow states [6,12], etc.

From a theoretical perspective, due to the presence of strong and subtle electron
correlation effects as well as an infinite number of open channels associated with these 
resonances, accurate and dependable calculation of their energies and widths have 
posed formidable difficulties to the theoreticians. This in addition with the availability
of numerous powerful and sophisticated quantum mechanical formalisms have stimulated an 
extensive amount of theoretical works in the recent years. In one of the earliest 
attempts, Ahmed and Lipsky [13] calculated the lowest triply excited resonances of Li-like 
atoms by a configuration-interaction (CI) type formalism. In a later work, Safronova and
Senashenko [14] performed a detailed and systematic study on the n=2 intrashell states of
Li isoelectronic series employing a 1/Z expansion approach for a wide range of Z. 
Thereafter, a multitude of theoretical methodologies have been proposed which provide
results in fairly good agreement with the experimental works, {\em viz.}, the joint saddle 
point (SP) and complex coordinate rotation (CCR) method [15-18], the 
state-specific theory [19,20], the space partition and stabilization approach [21],
the R-matrix theory along with many of its variants [22,23], the truncated 
diagonalization method (TDM) [24-26], the hyperspherical coordinate formalism [27,28], etc. 

While a number of formally attractive and elegant methodologies are now available within the 
traditional wave function based framework, attempts within a density functional viewpoint 
have been surprisingly very rare. The latter can be advantageous and desirable in the sense 
that the former often requires large basis sets and also possibly the mixing of continuum
states. Also whereas a substantial amount of experimental and theoretical results
exist for Li, only a few theoretical studies [17-19,21,25,26,29-33] and no experimental 
results exist as yet for the higher members of the isoelectronic series. Moreover, almost
all the publications on these ions have dealt exclusively only with the $2l2l'2l''$ intrashell
states; the only reported results on the $2l2l'nl'' (n>2)$ resonance series such as 
those considered in this work, are due to the recent TDM tabulations of Conneely 
{\em et al.} [25,26]. Furthermore to the best of my knowledge, we are not aware of any 
density-based results so far for the hollow resonances of these three-electron ions. The
objective of this article is therefore to provide results for such spectra in the lesser
explored Li-isoelectronic series using a density functional theory (DFT)-based approach 
with an aim to judge the performance of this formalism in comparison to the commonly 
used wave function based methodologies in the literature. Earlier an attempt 
was made to calculate all the 8 n=2 intrashell triply excited states of the 
Li-isoelectronic series (Z=2--4,6,8,10) [34] with reasonably good accuracy within the 
realm of nonrelativistic Hohenberg-Kohn-Sham (HKS) DFT. In this scheme, a local 
nonvariational work-function-based exchange potential [35] and the nonlinear 
correlation energy functional of Lee-Yang-Parr (LYP) [36] containing the gradient and 
Laplacian of the density was utilized. The resulting KS type differential equation was
numerically solved in the usual self-consistent manner exploiting the Numerov's type
finite difference (FD) scheme. Within a single-determinantal approach, this has 
shown to deliver fairly accurate results for various multiply excited states of closed-
and open-shell many-electron atoms and ions, e.g., the low and high as well as the
valence and core excitations, bound and autoionizing states, the satellite states, 
etc., ([37-42], and the references therein). However because of the Coulomb singularity
at r=0 and the long-range $-1/r$ behavior, the FD schemes for the radial discretization
often requires significantly larger number of grid points to achieve reasonable accuracy
even for the ground states and are not feasible for the high-lying Rydberg states. To
circumvent this problem, subsequently a generalized pseudospectral (GPS) technique was 
suggested for the numerical
solution of the KS equation for purposes of greater accuracy and efficiency [43]. 
Recently the GPS method has been applied quite successfully in the context of electronic 
structure and dynamics calculation of the Coulombic singularities including atoms, molecules
as well as other systems like the spiked harmonic oscillators with stronger singularities, 
the logarithmic potentials, etc [43-48]. Using this GPS implementation, the singly 
excited state energies for He, and Li and Be were calculated within 0.01\% and 0.02\% 
compared to the best available literature data; while the doubly excited states of He 
deviated by a maximum of 3.60\%. Furthermore the calculated single and double excitation 
energies were within the ranges of 0.009--0.632\% and 0.085--1.600\% respectively. This 
provided the motivation to apply this procedure to the more complicated and arguably more
interesting triply excited hollow resonance series of the three-electron atoms and ions. 
The validity and usefulness of this formalism has been amply demonstrated in a recent 
study [49] on the 12 $2l2l'$n$l''$ (n$\geq$2) hollow resonance series of Li covering a 
total of about 270 low- and high-lying states (with a maximum of n=25). The excited 
state energies were found to remain well within 0.98\% (with one exception) whereas the 
excitation energies showed discrepancies in the range of 0.021\%--0.058\%. Additionally 
37 $3l3l'$n$l''$ (n$\geq$3) doubly hollow states were reported, most of them for the 
first time ever.

Here we first compare the DFT excitation energies for all the 8 $2l2l'2l''$ 
triply excited states, {\em viz.,} 2s$^2$2p $^2$P$^o$; 2s2p$^2$ $^2$D$^e$, $^4$P$^e$, 
$^2$P$^e$, $^2$S$^e$; and 2p$^3$ $^2$D$^o$, $^2$P$^o$, $^4$S$^o$ of selected members 
of the Li-isoelectronic series, {\em viz.,} B$^{2+}$, N$^{4+}$ and F$^{6+}$. 
Then we focus on the 8 even- and odd-parity doublet and quartet $2l2l'$n$l''$ 
(3$\leq$n$\leq$6) hollow resonance series of all the members in the series; {\em viz.,} 
2s$^2$ns $^2$S$^e$, 2s$^2$np $^2$P$^o$, 2s$^2$nd $^2$D$^e$, 2s2pns $^4$P$^o$, 2s2pnp 
$^4$D$^e$, 2p$^2$ns $^4$P$^e$, 2p$^2$np $^4$D$^o$ and 2p$^2$ns $^2$D$^e$. In all the 
cases, comparisons with the available experimental and other theoretical results have been 
made, wherever possible. The article is organized as follows: Section II summarizes the
basic formalism as well as the computational strategy. Section III gives a discussion 
of the computed results while a few concluding remarks are presented in section IV. 

\section{Methodology} 
The calculation of atomic excited states within the work-function formalism [37-42,34] 
and its GPS implementation [43,49] has been presented in detail previously and will not
be repeated here. In the following therefore we present only a brief overview of it 
while addressing the essential problems associated with the excited state DFT and some 
recent developments in this direction. Atomic units are used throughout the article 
unless otherwise mentioned.

Ever since the formal inception in the works of Hohenberg-Kohn-Sham [50,51] nearly four
decades ago, DFT has emerged as a powerful practical tool for studying the ground-state
properties of many-electron systems such as atoms, molecules, solids etc. This is 
chiefly due to its ability to incorporate the electron correlation effects in a clever and
computationally inexpensive way. However inherent difficulties were recognized almost 
immediately for excited state calculations for the following reasons. No Hohenberg-Kohn like 
theorem exists for a general excited state as the wave function cannot be bypassed through 
the pure-state density in this case. This is easily realizable from the hydrodynamic 
formulation [52] of the wave function; the phase part of the hydrodynamic function is not 
constant for a general excited state, although it is constant for the ground and some static 
stationary excited states. In other words, the complex-valued wave function cannot be replaced 
by the real pure-state density function alone and consequently both the charge and current 
densities have to be taken into consideration. Also an accurate (not to speak of the exact) 
functional form of the exchange-correlation energy in terms of the density (incorporating the 
symmetry dependence) is as yet unknown and may not necessarily have the same dependence
on density as for the ground states. Thus although it is advantageous to work in terms of 
density for the ground states, it is not possible to characterize an individual excited state
solely in terms of density (angular momentum quantum numbers are required to classify them). 
Despite all these problems, a variety of promising and appealing formalisms have been 
suggested in the past years and this continues to remain a very fertile and demanding area of
research. Here we mention a few of the notable ones; others can be found in the review [42]. 
In the subspace formulation of DFT [53] for example, the intuitive idea of Slater's 
transition state theory was put into a rigorous foundation by invoking the concept of 
ensemble density instead 
of the pure-state density. Later, a formulation of the KS theory based on the constrained 
search method was developed [54] bypassing the Hohenberg-Kohn theorem and defining the 
functionals in terms of the Slater determinants. Several applications of the ensemble density
approach have been made in recent years [55,56] using a variety of approximate functionals. 
Further a Rayleigh-Ritz type variational principle of unequally weighted ensemble has been 
developed [57] as a generalization of the ensemble approach, where the lowest M eigenstates 
are weighted unequally. A time-dependent (TD) formulation of DFT for excited states [58-60], 
based on the TD density functional response theory, has shown considerable promise for 
excited states lately. The linear response of the fermionic system to a TD perturbation leads
to the frequency dependent dynamic dipole polarizability whose poles give the excitation 
energies. A perturbative treatment has been proposed as well [61,62], where the non-interacting
KS Hamiltonian corresponds to the zeroth-order Hamiltonian and the differences in the KS 
eigenvalues simply give the zeroth-order excitation energies. Recently multiplet energies 
for transition metal atoms, ions with $3d^n$ configurations have been reported [63] using the
SCF KS orbitals to construct the atomic multiplet determinantal combinations.

Despite all these attempts, a general prescription that can yield the bound and resonance 
states of a many-electron system reliably and accurately in a simple and uniform way  for
both low and high excitations, would be highly desirable. This is partly because, although 
some of the above methods produce good-quality results for certain states, cause significantly
large errors for other states making them unsuitable, presumably due to the fact that the 
potentials used are the ones obtained for the ground states and also importantly lacking the 
correct asymptotic behavior. Some others are difficult to implement computationally. 
Furthermore applications have been mostly restricted to the single excitations and low-lying 
states; results for double, triple or other multiple excitations as well as for higher lying 
states like the Rydberg series studied here, have been very scanty [43,49]. The work-function 
formulation of excited states as used in this work, either bypasses or partly overcomes the 
above mentioned problems of DFT in dealing with the excited states via an amalgamation of the 
conventional wave function based quantum mechanics within DFT such that the atomic orbital 
and the electronic configuration features are retained and we summarize its essential features
below. 

The point of central interest is the following KS type differential equation of the form,
\begin{equation}
\left[ -\frac{1}{2} \nabla^2 +\mathit{v}_{es} (\rvec) +\mathit{v}_{xc}(\rvec)
\right] \phi_i(\rvec) = \varepsilon_{\mathit{i}} \phi_{\mathit{i}}(\rvec),
\end{equation}
In this equation, $\mathit{v}_{es}(\rvec)$ signifies the usual Hartree electrostatic 
potential including the electron-nuclear attraction and the interelectronic Coulomb 
repulsion,
\begin{equation}
\mathit{v}_{es}(\rvec)=-\frac{Z}{r} + \int \frac{\rho(\rpvec)}{|\rvec-\rpvec|} 
\mathrm{d}\rpvec
\end{equation}
while $\mathit{v}_{xc}(\rvec)$, the total exchange-correlation (XC) potential is
partitioned as,
\begin{equation}
\mathit{v}_{xc}(\rvec)=\mathit{v}_{x}(\rvec)+\mathit{v}_{c}(\rvec)
\end{equation}
Now making the assumption that a unique exchange potential exists for a given excited state, 
one can physically interpret it as the the work done to move an electron against the electric 
field $\mbox{\boldmath $\cal{E}$}_x(\rvec)$ arising out of the Fermi-hole charge 
distribution, $\rho_x (\rvec,\rpvec)$, which can be expressed by the following line integral,
\begin{equation}
\mathit{v}_x (\rvec) = - \int_{\infty}^{r} 
\mbox{\boldmath $\cal{E}$}_x (\rvec) \cdot \mathrm{d} \mathbf{l}.
\end{equation}
where the field $\mbox{\boldmath $\cal{E}$}_x(\rvec)$ has the following form, 
\begin{equation} \label{eq:eqn1}
\mbox{\boldmath $\cal{E}$}_x(\rvec) = \int \frac 
{\rho_x (\rvec,\rpvec)(\rvec -\rpvec)} {|\rvec-\rpvec|^3} \ \ \mathrm{d}\rvec. 
\end{equation}
In order that the potential is well defined, the work done must be path-independent, 
(irrotational) and this is rigorously the case for spherically symmetric systems. This 
work against $\mbox{\boldmath $\cal{E}$}_x(\rvec)$ can be determined exactly as the 
Fermi hole is known explicitly in terms of the single-particle orbitals.
\begin{equation}
\rho_x(\rvec,\rpvec)=-\frac{|\gamma (\rvec,\rpvec)|^2}{2 \rho(\rvec)},
\end{equation}
where  
\begin{equation}
\gamma(\rvec,\rpvec)=\sum_{\mathit{i}} \phi_{\mathit{i}}^\ast(\rvec) 
\phi_{\mathit{i}}(\rpvec).
\end{equation}
denotes the single-particle density matrix spherically averaged over coordinates of 
the electrons of a given orbital angular momentum quantum number. Now within the 
central-field approximation, the orbitals expressed as $\phi_i(\rvec)= R_{nl}(r)\ 
Y_{lm}(\Omega)$, give the total electron density as the sum of the occupied orbitals,
\[\rho(\rvec)=\sum_{\mathit{i}} |\phi_{\mathit{i}}(\rvec)|^2. \]

At this point we note that the usual DFT route of obtaining the potential as a functional
derivative is avoided by introducing the physical interpretation that the dynamic 
Fermi-Coulomb hole charge determines the potential. The exchange potential does not have an
explicit density functional form (unlike the usual practice in DFT) and it is universal; 
the electronic configuration of a particular state defines it so that the same KS equation 
now equally valid for both ground and excited states. Although this formalism is in the 
spirit of Slater's interpretation of exchange energy based on the notion of Fermi hole, this
is expected to offer better results compared to the HFS scheme as it incorporates the dynamic 
nature of the charge distribution. The asymptotic behavior of the fully correlated potential 
(both $\mathit{v}_x(\rvec)$ and $\mathit{v}_c(\rvec)$) is that of $\mathit{v}_x(\rvec)$ only 
and falls off as $-1/r$ for atomic systems. However although this path allows an accurate 
determination of $\mathit{v}_{x}(\rvec)$ as elaborated, an analogous procedure leading to 
the accurate form of $\mathit{v}_{c}(\rvec)$ valid for a general excited state is as yet 
unknown, because the Coulomb hole is not known explicitly in terms of orbitals and therefore
requires to be approximated. Several approximate forms have been proposed in the literature 
ever since DFT's inception; this work uses the well-known LYP [36] functional which has been
found to be quite a good choice for the problem at hand. With this choice of 
$\mathit{v}_x(\rvec)$ and $\mathit{v}_c(\rvec)$, the KS equation is solved numerically to 
obtain the self-consistent set of orbitals and the densities therefrom. Because 
$\mathit{v}_x(\rvec)$ is local, the solution of the resultant KS equation is computationally 
much easier than the HF case. From this set of orbitals, various Slater determinants are 
constructed and different multiplet energies corresponding to a particular electronic 
configuration are calculated by following Slater's diagonal sum rule [64]. Various other 
aspects of this formalism can be found in the references [37-43,34,49].

Now a few words about the GPS discretization technique is in order. The commonly used 
equal-spacing FD schemes often require a significantly large number of spatial grid 
points because of the Coulomb singularity as well as the long-range nature of these 
physical systems; even for the accurate calculation of their ground states. The present 
Legendre pseudospectral discretization approach alleviates this uncomfortable feature in 
our calculation in three steps: (a) the semi-infinite domain $r \in [0, \infty]$ is mapped 
onto the finite domain $x \in [-1,1]$ by the transformation $r=r(x)$, (b) then an algebraic 
nonlinear mapping [65,66] is used, followed by a (c) symmetrization procedure leading to a 
set of coupled equations. The overall process gives rise to a symmetric matrix eigenvalue 
problem which can be accurately and easily solved to yield the eigenvalues and eigenfunctions 
by using the standard routines. This allows therefore nonuniform and optimal discretization, 
maintaining similar accuracies at both small (denser mesh) and large (coarser mesh) $r$ 
regions with a significantly smaller number of grid points with the promise of a fast 
convergence. We used a consistent convergence criteria of $10^{-6}$ and $10^{-8}$ a.u., for 
the potential and eigenvalues respectively; while a maximum of 500 radial grid points proved
to be sufficient to achieve convergence for all the states reported in this work.

\section{Results and discussion}
At first we present the computed density functional results for the nonrelativistic ground
state energies (in a.u.) of the Li-isoelectronic series, {\em viz.,} Be$^+$, B$^{2+}$, 
C$^{3+}$, N$^{4+}$, O$^{5+}$, F$^{6+}$ and Ne$^{7+}$ ions in Table I. These are compared
with the three best theoretical results in the literature, {\em viz.,} the fully 
correlated Hylleraas type variational method [67], the full core plus correlation using 
multiconfiguration interaction wave functions [68], and large expansion of the 
Hylleraas-type functions [69,70]. Our results show fairly good agreement with all of these. 
Slight overestimation is observed in the range of 0.0016\% for Be$^+$ to 0.0002\% for 
Ne$^{7+}$. Here we note that within the variationally founded HKS DFT, all many-body 
interactions are incorporated into the local multiplicative potential 
$\delta E_{xc}[\rho]/\delta \rho$, whose exact form remains unknown as yet. As stated earlier, 
the current formalism is not derived from the variational principle for energy. As a result, 
even though a KS type equation is solved to obtain the energies and densities, the total 
procedure is not subject to a variational bound. 

\begingroup
\squeezetable
\begin{table}
\caption {\label{tab:table1}Comparison of the calculated nonrelativistic ground-state
energies (in a.u.) of Li-isoelectronic series, Z=4--10 with the literature data. PW 
signifies present work.}
\begin{ruledtabular}
\begin{tabular}{ccc}
Z   & \multicolumn{2}{c}{Energy} \\ 
\cline{2-3}
   & PW  & Ref.  \\
\hline
4  & $-$14.32499  & $-$14.324763\footnotemark[1], $-$14.324761\footnotemark[2], 
                  $-$14.324760\footnotemark[3]  \\
5  & $-$23.42489  & $-$23.424605\footnotemark[1], $-$23.424603\footnotemark[2],
                  $-$23.424604\footnotemark[4]  \\
6  & $-$34.77586  & $-$34.775511\footnotemark[1], $-$34.775508\footnotemark[2],
                  $-$34.775509\footnotemark[4]  \\
7  & $-$48.37728  & $-$48.376898\footnotemark[1], $-$48.376895\footnotemark[2],
                  $-$48.376896\footnotemark[4]  \\
8  & $-$64.22893  & $-$64.228542\footnotemark[1], $-$64.228539\footnotemark[2],
                  $-$64.228540\footnotemark[4]  \\
9  & $-$82.33067  & $-$82.330338\footnotemark[1], $-$82.330335\footnotemark[2],
                  $-$82.330336\footnotemark[4]  \\
10 & $-$102.68243 & $-$102.682231\footnotemark[1], $-$102.682228\footnotemark[2],
                  $-$102.682229\footnotemark[4] \\
\end{tabular}
\end{ruledtabular}
\footnotetext[1]{Reference [67].} 
\footnotetext[2]{Reference [68].} 
\footnotetext[3]{Reference [69].} 
\footnotetext[4]{Reference [70].} 
\end{table}
\endgroup

\begingroup
\squeezetable
\begin{table}
\caption {\label{tab:table2}Comparison of the calculated excitation energies (in eV) of 
the n=2 intrashell triply excited states of B$^{2+}$, N$^{4+}$ and F$^{6+}$  relative 
to the nonrelativistic ground states of [67]. PW signifies present work.}
\begin{ruledtabular}
\begin{tabular}{lcccccc}
State & \multicolumn{2}{c}{B$^{2+}$} & \multicolumn{2}{c}{N$^{4+}$} & 
\multicolumn{2}{c}{F$^{6+}$} \\ 
\cline{2-3} \cline{4-5} \cline{6-7}
     &   PW  & Ref. &  PW  & Ref.  &  PW &  Ref. \\  
\hline
2s$^2$2p $^2$P$^o$   & 436.588  & 436.07\footnotemark[1],436.59\footnotemark[2]
                     & 894.541  & 893.93\footnotemark[1],894.12\footnotemark[3]
                     &1514.229  & 1515.67\footnotemark[1],1514.90\footnotemark[3] \\
2s2p$^2$ $^4$P$^e$   & 436.917  & 436.69\footnotemark[1],436.89\footnotemark[2]
                     & 894.876  & 894.54\footnotemark[1],894.51\footnotemark[3]
                     &1514.474  & 1516.33\footnotemark[1],1515.16\footnotemark[3]  \\
2s2p$^2$ $^2$D$^e$   & 441.893  & 441.34\footnotemark[1],442.00\footnotemark[2]
                     & 902.655  & 901.93\footnotemark[1],902.32\footnotemark[3]
                     &1524.898  & 1526.43\footnotemark[1],1525.86\footnotemark[3]  \\
2p$^3$ $^4$S$^o$     & 443.852  & 443.86\footnotemark[1],444.15\footnotemark[2],
                     & 905.329  & 905.15\footnotemark[1],905.15\footnotemark[3],
                     &1528.187  & 1530.42\footnotemark[1],1529.35\footnotemark[3],  \\
                     &          & 443.63\footnotemark[4]
                     &          & 904.43\footnotemark[4]
                     &          & 1528.51\footnotemark[4]                          \\
2s2p$^2$ $^2$S$^e$   & 445.387  & 445.11\footnotemark[1],445.75\footnotemark[2]
                     & 907.930  & 907.41\footnotemark[1],907.87\footnotemark[3]
                     &1531.822  & 1533.61\footnotemark[1],1533.18\footnotemark[3]  \\
2s2p$^2$ $^2$P$^e$   & 445.814  & 445.35\footnotemark[1],446.21\footnotemark[2]
                     & 908.455  & 907.99\footnotemark[1],908.59\footnotemark[3]
                     &1532.717  & 1534.55\footnotemark[1],1534.17\footnotemark[3]  \\
2p$^3$ $^2$D$^o$     & 446.173  & 446.02\footnotemark[1],446.58\footnotemark[2]
                     & 909.089  & 909.02\footnotemark[1],909.37\footnotemark[3]
                     &1533.816  & 1536.01\footnotemark[1],1535.35\footnotemark[3]  \\
2p$^3$ $^2$P$^o$     & 450.088  & 450.04\footnotemark[1],450.65\footnotemark[2]
                     & 915.023  & 915.00\footnotemark[1],915.47\footnotemark[3]
                     &1541.946  & 1543.95\footnotemark[1],1543.46\footnotemark[3]  \\
\end{tabular}
\end{ruledtabular}
\footnotetext[1]{Reference [29].}
\footnotetext[2]{Reference [25].}
\footnotetext[3]{Reference [26].}
\footnotetext[4]{Reference [71].}
\end{table}
\endgroup

Now Table II displays the computed nonrelativistic excitation energies in eV 
(1 a.u.=27.2076544 eV) of all the 8 triply excited states arising from electronic
configurations 2s$^2$2p, 2s2p$^2$ and 2p$^3$ of the ions B$^{2+}$, N$^{4+}$ and F$^{6+}$.
We have chosen these three ions partly because the work-function results have been previously 
reported (albeit within a different numerical procedure) for the even-Z series as well as
for He$^-$ and Li [34]; so we do not repeat them. Besides in the literature, relatively 
less results are available for the odd-Z series [25,26,29,71] than the even-Z series 
[17-19,25,26,29,71]. Therefore it may be useful to have the results for the odd-Z series, 
which will complete the work-function results for the isoelectronic series up to Z=10. Also
in this work, our primary focus in on the $2l2l'nl'' (n>2)$ resonance series. 
It is worth mentioning here that in our previous work on the hollow resonances in Li [49], 
the excited state energies were also presented in addition to the excitation energies. 
Throughout this work however, we do not report the state energies any more to avoid too 
many entries in the tables and also because that the available reference values of these
state energies for direct comparison are quite scarce. For a proper comparison, our density 
functional excitation energies presented in this and all other tables are estimated relative
to the respective accurate nonrelativistic ground states of these ions of [67]. No 
experimental results can be found for these states as yet and only a few theoretical
studies have been made. The 2p$^3$ $^4$S$^o$ states of these ions are bound, metastable
against autoionization by conservation of parity and angular momentum and were reported
by using a multiconfiguration-interaction type formalism within a Rayleigh-Ritz 
variational principle [71]. Recently the position of all these states for these ions have 
been calculated by Safronova and Bruch [29] on the basis of a perturbation theory 
method (1/Z expansion) as well as by Conneely {\em et al.} [25,26] following the TDM. 
It is evident that the current positions of these states for all the three ions follow the
same rank orderings of [29] and [25,26]. For all these states, excellent agreement with the
literature data is noticed with the largest absolute discrepancy of 0.125\% (the deviation 
ranges from 0.0005\%--0.125\%, 0.007\%--0.049\% and 0.044\%--0.099\% for the three ions 
respectively). The present excitation energies for B$^{2+}$ show underestimations for all
but the 2s2p$^2$ $^4$P$^e$ state compared to [25] and overestimations for all but the 2p$^3$ 
$^4$S$^o$ state relative to [29]. N$^{4+}$ gives overestimations for all the states relative
to [29], while overestimations for all but the last three states with respect to [26]. 
F$^{6+}$ excitation energies are underestimated for all the states compared to all the 
literature results. Note that excitation energies of [25,26] and [71] are quoted here with 
respect to the same ground states of [67]. 

Now we turn our focus on to the $2l2l'nl'' (n\geq 3)$ hollow resonance series. The excitation
energies (in eV) for the 8 such even- and odd-parity series, {\em viz.,} 
$\langle$A,ns$\rangle$ $^2$S$^e$, $\langle$A,np$\rangle$ $^2$P$^o$,  
$\langle$A,nd$\rangle$ $^2$D$^e$, $\langle$B,ns$\rangle$ $^4$P$^o$,
$\langle$B,np$\rangle$ $^4$D$^e$, $\langle$C,ns$\rangle$ $^4$P$^e$ and
$\langle$C,np$\rangle$ $^4$D$^o$, $\langle$D,ns$\rangle$ $^2$D$^e$ of the Li-like ions with
Z=4--10, relative to the same ground states [67] as used in Table II, are tabulated in 
Tables III and IV respectively. In these two tables we restrict ourselves to the lower 
resonances with $3\leq n \leq 6$, whereas the higher resonances are given later. The
independent particle model classification of [24-26] has been utilized, i.e., these states
are classified according to the Rydberg series which converge on each of the six lowest 
$2l2l'$ doubly excited core states of the residual two-electron systems, {\em viz.,} 
2s$^2$ $^1$S$^e$, 2s2p $^3$P$^o$, 2p$^2$ $^3$P$^e$, 2p$^2$ $^1$D$^e$, 2s2p $^1$P$^o$ and 
2p$^2$ $^1$S$^e$, denoted by A, B, C, D, E and F respectively. Again no experimental results
can be found for any of these resonances and we are not aware of any other theoretical 
attempts except for the recent extensive TDM tabulations [25,26]. We quote the TDM excitation 
energies relative to the same ground states of [67] as used for the present calculations. At
a first glance, the overall agreement of the density functional excitation energies with the
reference values is quite gratifying. The largest absolute deviation is found to be 0.57\% 
for the $\langle$A,3p$\rangle$ $^2$P$^o$ state of Be$^+$; for all other cases, the same 
remains well below 0.50\%. In 3 instances, our results completely match with the TDM values, 
e.g., the $\langle$A,3s$\rangle$ $^2$S$^e$ of O$^{5+}$, $\langle$A,3p$\rangle$ $^2$P$^o$
and $\langle$C,3s$\rangle$ $^4$P$^e$ states of Ne$^{7+}$. The absolute deviation ranges are
0.042\%--0.566\%, 0.016\%--0.375\%, 0.007\%--0.247\%, 0.004\%--0.155\%, 0.000\%--0.102\%, 
0.002\%--0.090\% and 0.000\%--0.103\% for Z=4,5,6,7,8,9 and 10 ions respectively. As in the 
previous table, here also we notice both over- and under-estimations. Generally speaking, 
accuracy of our calculation increases with an increase in Z for a particular state. This is
partly due to the fact that electrons are pulled closer to the nucleus with an increase in
Z; the depth of the Coulomb potential increases and the one-electron contributions to the 
energy becomes more important than the correlation term (which is dealt with comparatively
less accurately).

\begingroup
\squeezetable
\begin{table}
\caption {\label{tab:table3}Calculated excitation energies (in eV) of various triply 
excited $2l2l'nl'' (3\leq n \leq 6)$ resonances  of Be$^+$, B$^{2+}$ and C$^{3+}$ 
relative to the nonrelativistic ground states of [67]. PW signifies present work and the 
TDM values are quoted from Ref. [25]. }
\begin{ruledtabular}
\begin{tabular}{lllllll}
State & \multicolumn{2}{c}{Be$^+$} & \multicolumn{2}{c}{B$^{2+}$} & 
\multicolumn{2} {c}{C$^{3+}$} \\
\cline{2-3} \cline{4-5} \cline{6-7}
     &   PW  & TDM &  PW  & TDM & PW & TDM \\  
\hline
$\langle$A,3s$\rangle$ $^2$S$^e$   & 285.924  & 284.694 
                                   & 466.320  & 465.150  & 691.143 & 690.220  \\
$\langle$A,4s$\rangle$ $^2$S$^e$   & 290.424  & 289.192 
                                   & 475.138  & 473.495  & 705.606 & 705.114  \\
$\langle$A,5s$\rangle$ $^2$S$^e$   & 292.217  & 291.028 
                                   & 478.795  & 477.916  & 711.742 & 711.083  \\
$\langle$A,6s$\rangle$ $^2$S$^e$   & 293.118  & 292.103 
                                   & 480.670  & 479.647  & 714.928 & 714.190  \\
$\langle$A,3p$\rangle$ $^2$P$^o$   & 287.034  & 285.418 
                                   & 467.904  & 466.157  & 693.205 & 691.499  \\
$\langle$A,4p$\rangle$ $^2$P$^o$   & 290.827  & 289.570 
                                   & 475.726  & 474.583  & 706.395 & 705.965  \\
$\langle$A,5p$\rangle$ $^2$P$^o$   & 292.405  & 291.325 
                                   & 479.075  & 478.139  & 712.136 & 711.426  \\
$\langle$A,6p$\rangle$ $^2$P$^o$   & 293.221  & 292.111 
                                   & 480.830  & 479.872  & 715.156 & 714.479  \\
$\langle$A,3d$\rangle$ $^2$D$^e$   & 288.563  & 287.355 
                                   & 470.116  & 469.022  & 696.092 & 695.265  \\
$\langle$A,4d$\rangle$ $^2$D$^e$   & 291.371  & 290.144 
                                   & 476.520  & 475.476  & 707.462 & 706.673  \\
$\langle$A,5d$\rangle$ $^2$D$^e$   & 292.672  & 291.556 
                                   & 479.467  & 478.477  & 712.672 & 711.908  \\
$\langle$A,6d$\rangle$ $^2$D$^e$   & 293.382  & 292.285 
                                   & 481.056  & 480.066  & 715.480 & 714.740  \\
$\langle$B,3s$\rangle$ $^4$P$^o$   & 285.331  & 285.508 
                                   & 466.081  & 466.184  & 691.515 & 691.464  \\
$\langle$B,4s$\rangle$ $^4$P$^o$   & 289.867  & 290.468 
                                   & 474.951  & 475.367  & 706.042 & 706.374  \\
$\langle$B,5s$\rangle$ $^4$P$^o$   & 291.662  & 292.152 
                                   & 478.613  & 479.056  & 712.185 & 712.778  \\
$\langle$B,6s$\rangle$ $^4$P$^o$   & 292.566  & 293.058 
                                   & 480.487  & 481.012  & 715.374 & 715.806  \\
$\langle$B,3p$\rangle$ $^4$D$^e$   & 286.120  & 286.400 
                                   & 467.196  & 467.468  & 692.941 & 693.143  \\
$\langle$B,4p$\rangle$ $^4$D$^e$   & 290.169  & 290.495 
                                   & 475.400  & 475.867  & 706.629 & 707.024  \\
$\langle$B,5p$\rangle$ $^4$D$^e$   & 291.804  & 292.318 
                                   & 478.830  & 479.320  & 712.474 & 712.748  \\
$\langle$B,6p$\rangle$ $^4$D$^e$   & 292.642  & 293.148 
                                   & 480.612  & 481.157  & 715.532 & 715.989  \\
$\langle$C,3s$\rangle$ $^4$P$^e$   & 288.792  & 289.284 
                                   & 470.570  & 471.152  & 697.025 & 697.621  \\
$\langle$C,4s$\rangle$ $^4$P$^e$   & 293.058  & 293.567 
                                   & 479.040  & 479.647  & 711.023 & 711.606  \\
$\langle$C,5s$\rangle$ $^4$P$^e$   & 294.788  & 295.417 
                                   & 482.601  & 483.252  & 717.033 & 717.648  \\
$\langle$C,6s$\rangle$ $^4$P$^e$   & 295.667  & 296.288 
                                   & 484.441  & 485.113  & 720.171 & 720.813  \\
$\langle$C,3p$\rangle$ $^4$D$^o$   & 289.483  & 290.011 
                                   & 471.601  & 472.279  & 698.440 & 699.131  \\
$\langle$C,4p$\rangle$ $^4$D$^o$   & 293.477  & 293.833 
                                   & 479.693  & 480.033  & 711.962 & 712.079  \\
$\langle$C,5p$\rangle$ $^4$D$^o$   & 295.099  & 294.545 
                                   & 483.097  & 483.458  & 717.760 & 717.877  \\
$\langle$C,6p$\rangle$ $^4$D$^o$   & 295.926  & 296.358 
                                   & 484.862  & 485.216  & 720.800 & 720.943  \\
$\langle$D,3s$\rangle$ $^2$D$^e$   & 290.710  & 290.264 
                                   & 472.959  & 472.883  & 698.701 & 700.031  \\
$\langle$D,4s$\rangle$ $^2$D$^e$   & 294.857  & 294.982 
                                   & 481.249  & 481.636  & 712.803 & 714.103  \\
$\langle$D,5s$\rangle$ $^2$D$^e$   & 296.557  & 296.979 
                                   & 484.767  & 485.349  & 719.039 & 720.372  \\
$\langle$D,6s$\rangle$ $^2$D$^e$   & 297.425  & 297.868 
                                   & 486.590  & 487.281  & 722.236 & 723.593  \\
\end{tabular}
\end{ruledtabular}
\end{table}
\endgroup

\begingroup
\squeezetable
\begin{table}
\caption {\label{tab:table4}Calculated excitation energies (in eV) of various triply 
excited $2l2l'nl'' (3 \leq n \leq 6)$ resonances  of N$^{4+}$, O$^{5+}$, F$^{6+}$ and 
Ne$^{7+}$ relative to the nonrelativistic ground states of [67]. PW signifies present 
work and the TDM values are taken from Ref. [26].}
\begin{ruledtabular}
\begin{tabular}{lllllllll}
State & \multicolumn{2}{c}{N$^{4+}$} & \multicolumn{2}{c}{O$^{5+}$}
      & \multicolumn{2}{c}{F$^{6+}$} & \multicolumn{2}{c}{Ne$^{7+}$}  \\
\cline{2-3} \cline{4-5} \cline{6-7} \cline{8-9}
     &   PW  & TDM  &  PW  & TDM  &  PW  &   TDM  &  PW  &  TDM   \\
\hline
$\langle$A,3s$\rangle$ $^2$S$^e$ & 960.417  &  959.883  & 1274.141 & 1274.141  
                                 & 1632.322 &  1632.988 & 2035.060 & 2036.423     \\
$\langle$A,4s$\rangle$ $^2$S$^e$ & 981.845  &  981.747  & 1303.860 & 1304.322  
                                 & 1671.653 &  1672.828 & 2085.326 & 2087.250     \\
$\langle$A,5s$\rangle$ $^2$S$^e$ & 991.077  &  990.786  & 1316.797 & 1316.911  
                                 & 1688.908 &  1690.421 & 2107.514 & 2109.696     \\
$\langle$A,6s$\rangle$ $^2$S$^e$ & 995.903  &  995.697  & 1323.602 & 1324.026  
                                 & 1698.023 &  1699.146 & 2119.270 & 2121.150     \\
$\langle$A,3p$\rangle$ $^2$P$^o$ & 962.928  &  961.437  & 1277.093 & 1275.961  
                                 & 1635.682 &  1635.078 & 2038.779 & 2038.779     \\
$\langle$A,4p$\rangle$ $^2$P$^o$ & 982.814  &  982.555  & 1305.000 & 1305.294  
                                 & 1672.934 &  1673.966 & 2086.700 & 2088.553     \\
$\langle$A,5p$\rangle$ $^2$P$^o$ & 991.556  &  991.398  & 1317.360 & 1317.714  
                                 & 1689.523 &  1690.453 & 2108.129 & 2110.305     \\
$\langle$A,6p$\rangle$ $^2$P$^o$ & 996.184  &  996.146  & 1323.923 & 1324.315  
                                 & 1698.357 &  1699.505 & 2119.564 & 2121.531     \\
$\langle$A,3d$\rangle$ $^2$D$^e$ & 966.440  &  966.089  & 1281.261 & 1281.495  
                                 & 1640.468 &  1641.485 & 2044.120 & 2046.063     \\
$\langle$A,4d$\rangle$ $^2$D$^e$ & 984.120  &  983.804  & 1306.586 & 1306.973  
                                 & 1674.760 &  1675.867 & 2088.708 & 2090.746     \\
$\langle$A,5d$\rangle$ $^2$D$^e$ & 992.211  &  991.907  & 1318.168 & 1318.441  
                                 & 1690.451 &  1691.523 & 2109.116 & 2111.111     \\
$\langle$A,6d$\rangle$ $^2$D$^e$ & 996.570  &  996.279  & 1324.413 & 1324.696  
                                 & 1698.912 &  1699.973 & 2120.127 & 2122.026     \\
$\langle$B,3s$\rangle$ $^4$P$^o$ & 961.584  &  961.336  & 1276.318 & 1275.798  
                                 & 1635.695 &  1634.849 & 2039.772 & 2038.488     \\
$\langle$B,4s$\rangle$ $^4$P$^o$ & 983.089  &  983.306  & 1306.126 & 1306.154  
                                 & 1675.130 &  1674.912 & 2090.153 & 2089.581     \\
$\langle$B,5s$\rangle$ $^4$P$^o$ & 992.328  &  992.696  & 1319.074 & 1319.240  
                                 & 1692.399 &  1692.325 & 2112.357 & 2111.938     \\
$\langle$B,6s$\rangle$ $^4$P$^o$ & 997.160  &  997.471  & 1325.882 & 1325.917  
                                 & 1701.516 &  1701.603 & 2124.113 & 2123.806     \\
$\langle$B,3p$\rangle$ $^4$D$^e$ & 963.339  &  963.409  & 1278.410 & 1278.268  
                                 & 1638.076 &  1637.714 & 2042.621 & 2041.748     \\
$\langle$B,4p$\rangle$ $^4$D$^e$ & 983.837  &  984.125  & 1307.043 & 1307.144  
                                 & 1676.161 &  1676.074 & 2091.486 & 2090.917     \\
$\langle$B,5p$\rangle$ $^4$D$^e$ & 992.707  &  993.093  & 1319.553 & 1319.722  
                                 & 1692.927 &  1692.897 & 2113.113 & 2112.599     \\
$\langle$B,6p$\rangle$ $^4$D$^e$ & 997.378  &  997.713  & 1326.170 & 1326.279  
                                 & 1701.821 &  1701.902 & 2124.619 & 2124.154     \\
$\langle$C,3s$\rangle$ $^4$P$^e$ & 968.130  &  968.685  & 1283.917 & 1284.333  
                                 & 1644.334 &  1644.573 & 2049.398 & 2049.398     \\
$\langle$C,4s$\rangle$ $^4$P$^e$ & 988.982  &  989.515  & 1312.945 & 1313.336  
                                 & 1682.860 &  1683.064 & 2098.745 & 2098.704     \\
$\langle$C,5s$\rangle$ $^4$P$^e$ & 998.053  &  998.600  & 1325.691 & 1326.034  
                                 & 1699.892 &  1700.044 & 2120.677 & 2120.568     \\
$\langle$C,6s$\rangle$ $^4$P$^e$ & 1002.817 &  1003.375 & 1332.420 & 1332.852  
                                 & 1708.917 &  1709.118 & 2132.330 & 2132.273     \\
$\langle$C,3p$\rangle$ $^4$D$^o$ & 969.947  &  970.570  & 1286.183 & 1286.597  
                                 & 1647.090 &  1647.210 & 2052.633 & 2052.410     \\
$\langle$C,4p$\rangle$ $^4$D$^o$ & 990.223  &  990.119  & 1314.536 & 1314.066  
                                 & 1684.843 &  1683.924 & 2101.107 & 2099.689     \\
$\langle$C,5p$\rangle$ $^4$D$^o$ & 999.032  &  998.221  & 1326.970 & 1326.399  
                                 & 1701.502 &  1700.488 & 2122.622 & 2121.077     \\
$\langle$C,6p$\rangle$ $^4$D$^o$ & 1003.677 &  1003.579 & 1333.552 & 1333.048  
                                 & 1710.353 &  1709.357 & 2134.077 & 2132.556     \\
$\langle$D,3s$\rangle$ $^2$D$^e$ & 971.302  &  971.765  & 1287.386 & 1288.082  
                                 & 1648.072 &  1648.981 & 2053.360 & 2054.464     \\
$\langle$D,4s$\rangle$ $^2$D$^e$ & 991.863  &  992.143  & 1316.065 & 1316.626  
                                 & 1686.193 &  1686.955 & 2102.247 & 2103.166     \\
$\langle$D,5s$\rangle$ $^2$D$^e$ & 1000.861 &  1001.816 & 1328.720 & 1329.822  
                                 & 1703.121 &  1704.441 & 2124.056 & 2125.602     \\
$\langle$D,6s$\rangle$ $^2$D$^e$ & 1005.595 &  1006.770 & 1335.413 & 1336.776  
                                 & 1712.103 &  1713.653 & 2135.663 & 2137.382     \\
\end{tabular}
\end{ruledtabular}
\end{table} 
\endgroup

To further demonstrate the usefulness and applicability of the prescription, we now report
the excitation energies (relative to the same ground state of [67]) corresponding to the 
high-lying Rydberg states of the 2s$^2$ns $^2$S$^e$ (n=7--25) hollow resonances for all 
the 7 ionic species. To the best of my knowledge, no experimental observation has been made
as of today and no other theoretical results exist for any of these resonances except the 
recent TDM calculations [25,26], which are appropriately quoted. These reference results 
are available for resonances up to n=14, 15, 17 for Be$^+$, B$^{2+}$, F$^{6+}$ and n=16 for
the remaining ions. The present density functional excitation energies are overestimated
for Be$^+$, B$^{2+}$, C$^{3+}$, N$^{4+}$ with mean absolute deviations 0.363\%, 0.205\%,
0.098\%, 0.028\% respectively, while the same for O$^{5+}$, F$^{6+}$, Ne$^{7+}$ show the 
opposite trend with the mean absolute deviations 0.020\%, 0.057\% and 0.080\% respectively.
Once again the deviations are larger for low Z and tends to diminish with an increase in Z. 
As pointed out in several recent works for the hollow Rydberg series of Li including the 
present formalism [22-24,49], these resonances are also expected to be highly entangled to 
each other, making their precise theoretical calculation as well as experimental observation 
significantly complicated. However, as expected, the separation between the successive 
members within a resonance series increases as Z goes up. The resonances above n=17 are 
reported in this work for the first time. At this stage it is worthwhile to mention that 
although the TDM method is advantageous in the sense of obtaining the whole resonance series 
at once as well as enabling an accurate classification of the energy levels on the basis of 
configuration mixing and quantum defects in a uniform and consistent manner, the excitation 
energies are generally not as accurate as some of the other existing approaches (such as the 
saddle point method) [24-26]. Thus more precise elaborate calculational and experimental 
results would be needed to resolve some of the discrepancies observed in this work as well 
as to unveil various interesting features associated with these resonances. Finally, in 
Fig. 1 we show the radial densities of Be$^+$ for 2s$^2$3p $^2$P$^o$, 2p$^2$3s $^2$D$^e$, 
2s2p3s $^4$P$^o$ in (a) and 2s$^2$3s $^2$S$^e$, 2s$^2$4s$^2$S$^e$ in (b) while Fig. 2 depicts 
the radial densities of 2s$^2$3s $^2$S$^e$ states for B$^{2+}$, C$^{3+}$, N$^{4+}$ in (a) 
and for O$^{5+}$, F$^{6+}$, Ne$^{7+}$ in (b) respectively. As expected, the peaks are
shifted to smaller values of r as Z increases and of course the desired shell structure is 
observed in all these plots. 

\begingroup
\squeezetable
\begin{table}
\caption {\label{tab:table5}Calculated excitation energies (in eV) of the 2s$^2$ns
$^2$S$^e$ (n=7--25) resonances  of the Li-isoelectronic series (Z=4--10) relative to
the nonrelativistic ground states of [67]. PW signifies present work and the TDM 
values are quoted from Refs. [25] and [26].}
\begin{ruledtabular}
\begin{tabular}{lllllllll}
n     & \multicolumn{2}{c}{Be$^+$}   & \multicolumn{2}{c}{B$^{2+}$} & 
\multicolumn{2}{c}{C$^{3+}$} & \multicolumn{2}{c}{N$^{4+}$}      \\
\cline{2-3} \cline{4-5} \cline{6-7} \cline{8-9}
   & PW      & TDM     & PW      & TDM     & PW      & TDM     & PW       &  TDM     \\  
\hline
7  & 293.635 & 292.636 & 481.761 & 480.770 & 716.794 & 716.117 & 998.747  & 998.543  \\
8  & 293.959 & 292.900 & 482.452 & 481.486 & 717.980 & 717.311 & 1000.561 & 1000.295 \\
9  & 294.176 & 293.107 & 482.917 & 481.927 & 718.780 & 718.100 & 1001.789 & 1001.500 \\
10 & 294.329 & 293.249 & 483.243 & 482.240 & 719.346 & 718.644 & 1002.659 & 1002.311 \\
11 & 294.440 & 293.368 & 483.483 & 482.446 & 719.762 & 719.069 & 1003.296 & 1003.035 \\
12 & 294.522 & 293.447 & 483.662 & 482.694 & 720.075 & 719.363 & 1003.777 & 1003.560 \\
13 & 294.587 & 293.510 & 483.801 & 482.822 & 720.318 & 719.599 & 1004.153 & 1003.856 \\
14 & 294.636 & 293.561 & 483.910 & 482.941 & 720.508 & 719.795 & 1004.447 & 1004.139 \\
15 & 294.677 &         & 483.999 & 483.018 & 720.663 & 719.942 & 1004.683 & 1004.390 \\
16 & 294.710 &         & 484.071 &         & 720.788 & 720.064 & 1004.877 & 1004.572 \\
17 & 294.737 &         & 484.130 &         & 720.892 &         & 1005.037 &          \\
18 & 294.759 &         & 484.179 &         & 720.976 &         & 1005.170 &          \\
19 & 294.778 &         & 484.220 &         & 721.049 &         & 1005.285 &          \\
20 & 294.794 &         & 484.256 &         & 721.112 &         & 1005.380 &          \\
21 & 294.808 &         & 484.286 &         & 721.166 &         & 1005.462 &          \\
22 & 294.821 &         & 484.313 &         & 721.213 &         & 1005.532 &          \\ 
23 & 294.832 &         & 484.334 &         & 721.251 &         & 1005.595 &          \\
24 & 294.843 &         & 484.356 &         & 721.286 &         & 1005.649 &          \\
25 & 294.851 &         & 484.373 &         & 721.319 &         & 1005.698 &          \\
\hline
      & \multicolumn{2}{c}{O$^{5+}$}   & \multicolumn{2}{c}{F$^{6+}$} & 
\multicolumn{2}{c}{Ne$^{7+}$} &        &      \\
\hline
7  & 1327.620 & 1327.958 & 1703.421 & 1704.441 & 2126.246 & 2128.072 &     &    \\
8  & 1330.192 & 1330.450 & 1706.879 & 1708.005 & 2130.722 & 2132.542 &     &    \\
9  & 1331.938 & 1332.156 & 1709.227 & 1710.247 & 2133.767 & 2135.489 &     &    \\
10 & 1333.173 & 1333.478 & 1710.894 & 1711.893 & 2135.927 & 2137.641 &     &    \\
11 & 1334.082 & 1334.360 & 1712.122 & 1713.101 & 2137.521 & 2139.170 &     &    \\
12 & 1334.771 & 1335.040 & 1713.052 & 1714.007 & 2138.727 & 2140.432 &     &    \\
13 & 1335.301 & 1335.560 & 1713.770 & 1714.739 & 2139.660 & 2141.360 &     &    \\
14 & 1335.723 & 1335.979 & 1714.339 & 1715.267 & 2140.400 & 2142.098 &     &    \\
15 & 1336.063 & 1336.311 & 1714.799 & 1715.770 & 2140.996 & 2142.674 &     &    \\
16 & 1336.338 & 1336.591 & 1715.172 & 1716.121 & 2141.480 & 2143.148 &     &    \\
17 & 1336.566 &          & 1715.479 & 1716.420 & 2141.883 &          &     &    \\
18 & 1336.757 &          & 1715.737 &          & 2142.220 &          &     &    \\
19 & 1336.917 &          & 1715.958 &          & 2142.503 &          &     &    \\
20 & 1337.056 &          & 1716.143 &          & 2142.745 &          &     &    \\
21 & 1337.173 &          & 1716.303 &          & 2142.955 &          &     &    \\
22 & 1337.276 &          & 1716.442 &          & 2143.134 &          &     &    \\
23 & 1337.366 &          & 1716.562 &          & 2143.289 &          &     &    \\
24 & 1337.442 &          & 1716.668 &          & 2143.428 &          &     &    \\
25 & 1337.513 &          & 1716.760 &          & 2143.550 &          &     &    \\
\end{tabular}
\end{ruledtabular}
\end{table} 
\endgroup

\begin{figure}
\centering
\begin{minipage}[c]{0.40\textwidth}
\centering
\includegraphics[scale=0.45]{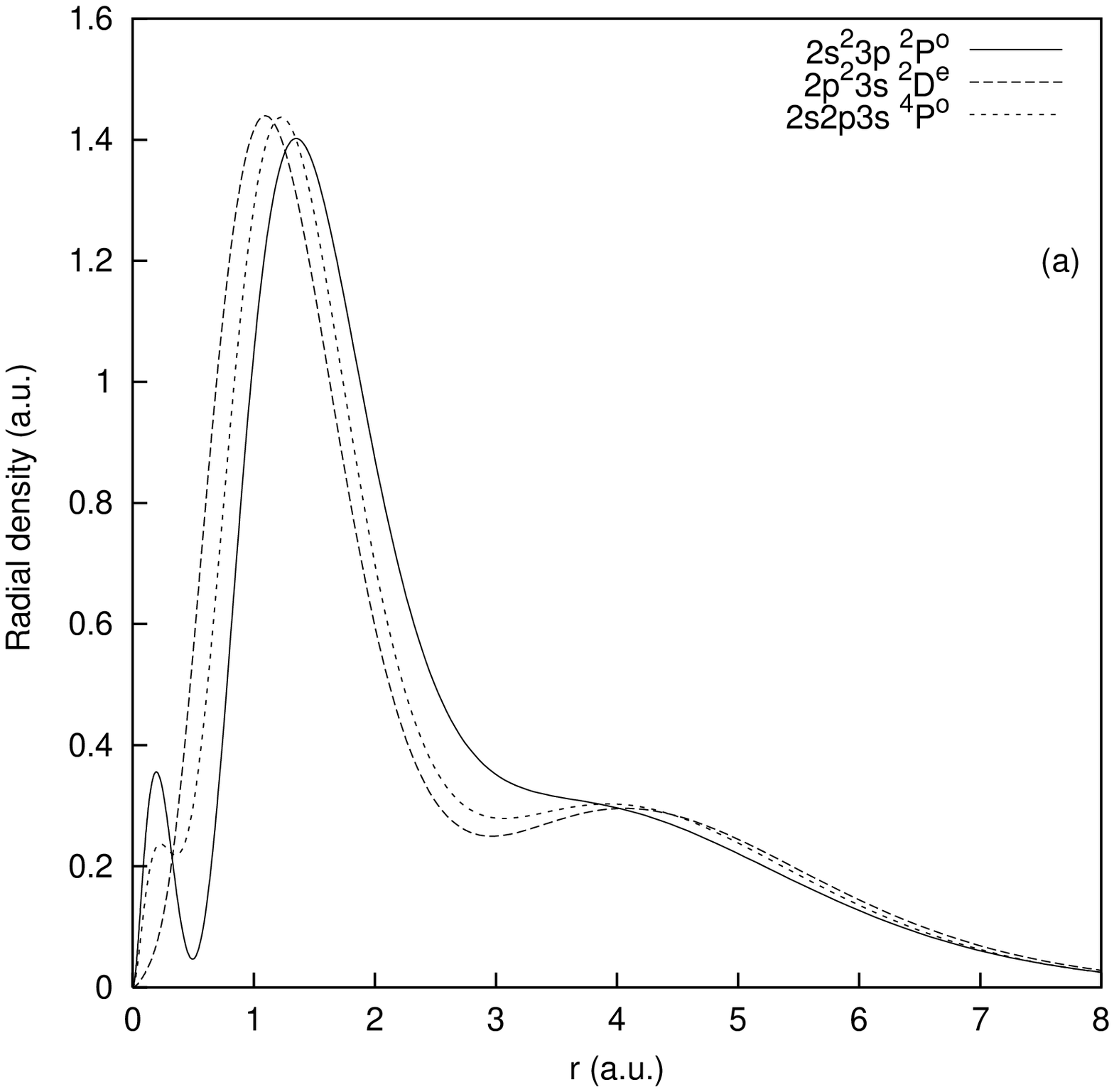}
\end{minipage}%
\hspace{0.6in}
\begin{minipage}[c]{0.40\textwidth}
\centering
\includegraphics[scale=0.45]{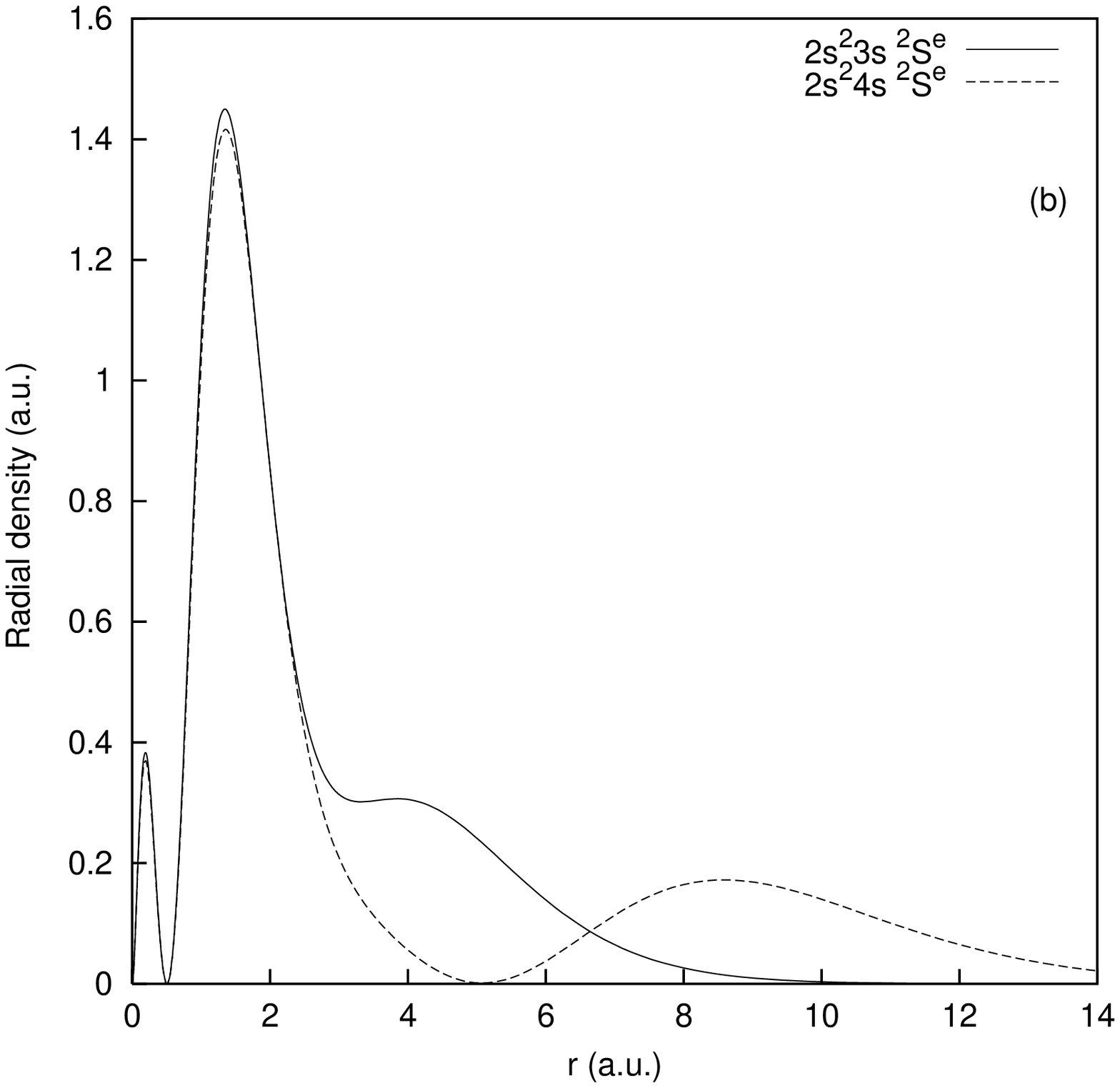}
\end{minipage}%

\caption{The radial densities (a.u.) of Be$^+$ for (a) 2s$^2$3p $^2$P$^o$, 2p$^2$3s 
$^2$D$^e$, 2s2p3s $^4$P$^o$ and (b) 2s$^2$3s $^2$S$^e$, 2s$^2$4s $^2$S$^e$.}
\end{figure}

\begin{figure}
\centering
\begin{minipage}[c]{0.40\textwidth}
\centering
\includegraphics[scale=0.45]{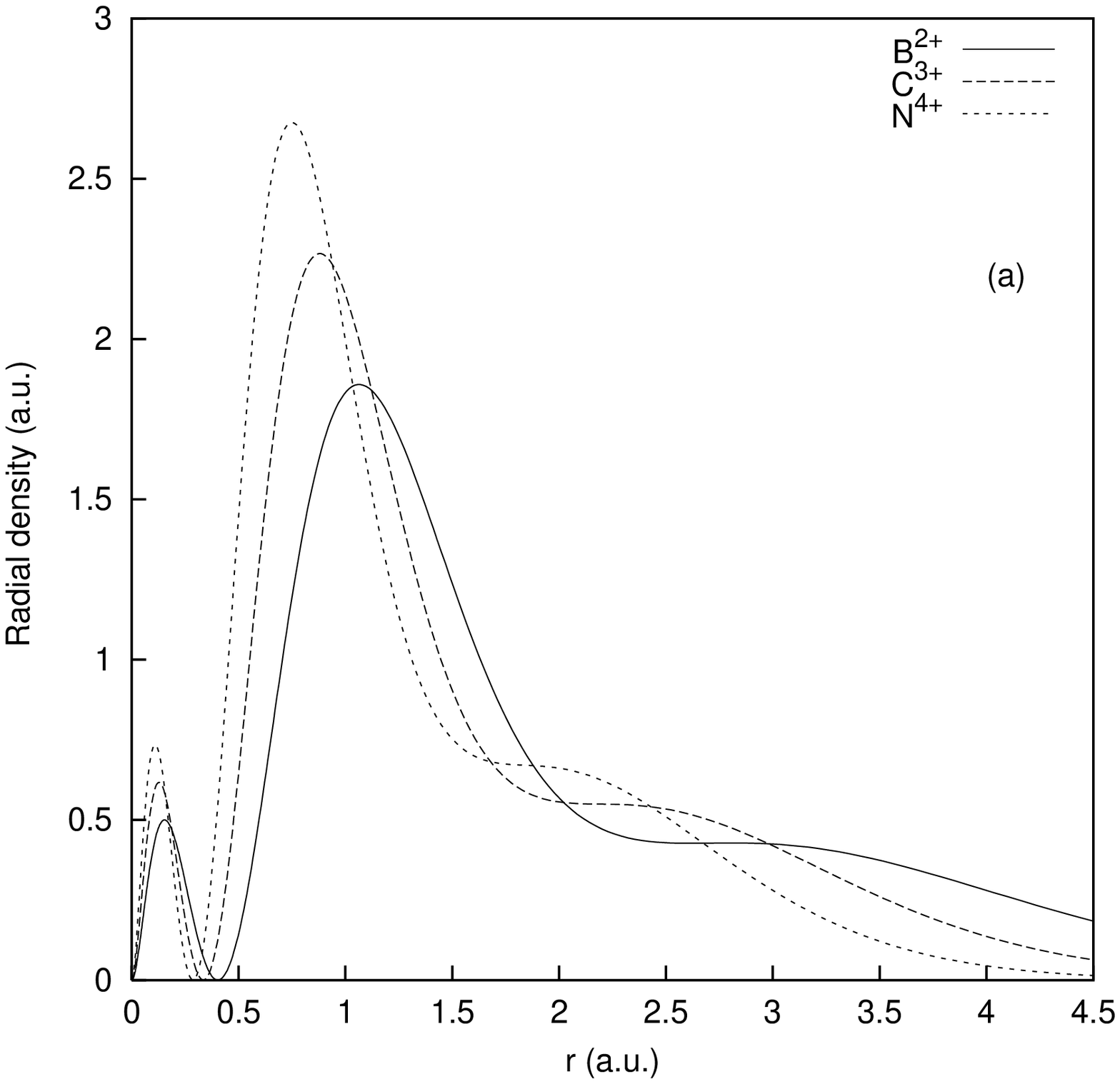}
\end{minipage}%
\hspace{0.6in}
\begin{minipage}[c]{0.40\textwidth}
\centering
\includegraphics[scale=0.45]{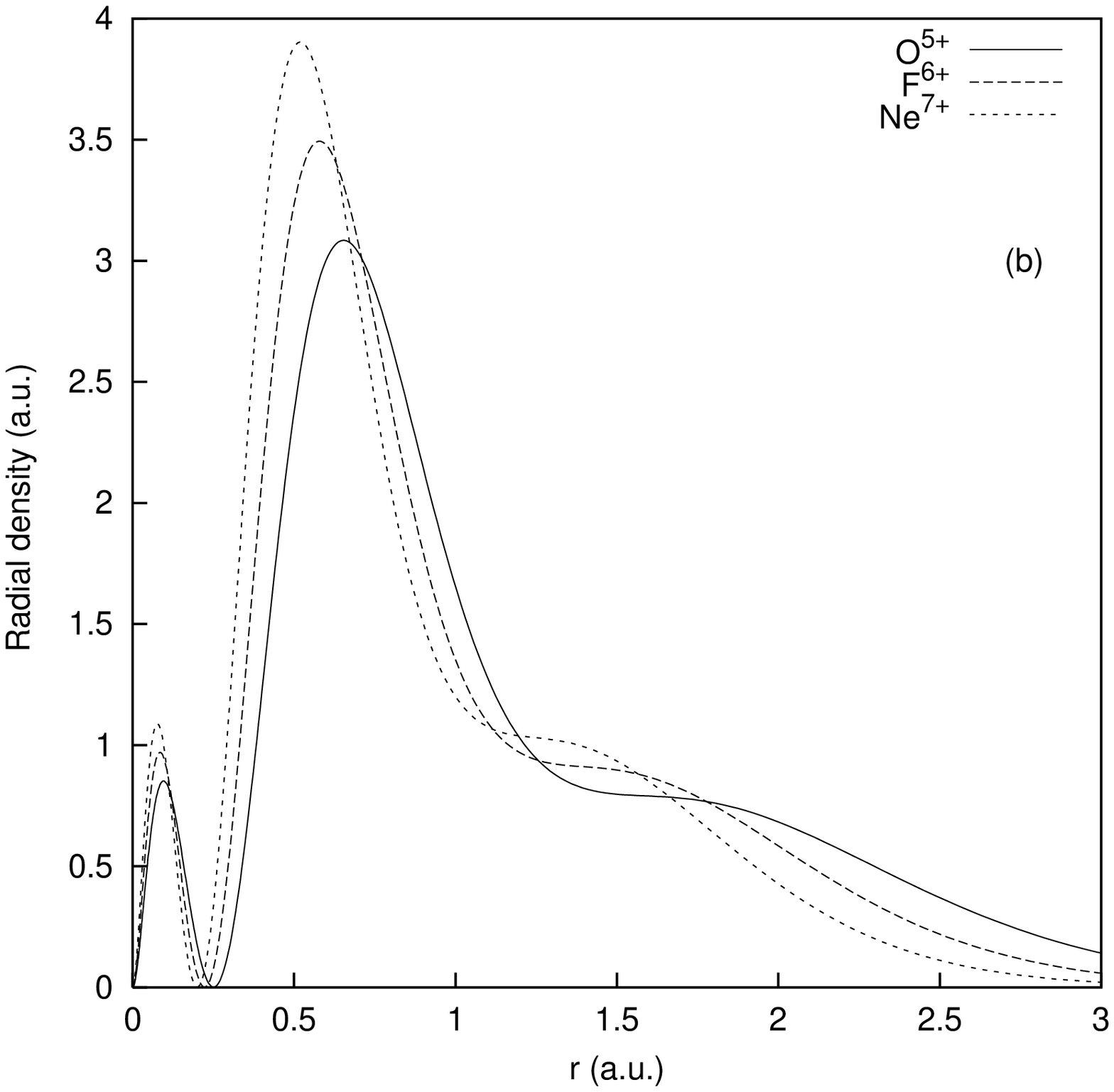}
\end{minipage}%

\caption{The radial densities (a.u.) of 2s$^2$3s $^2$S$^e$ states for (a) B$^{2+}$,
C$^{3+}$, N$^{4+}$ and (b) O$^{5+}$, F$^{6+}$, Ne$^{7+}$ respectively.}
\end{figure}

\section{Conclusion}
In this follow-up of our proposed density functional approach to the triply excited states 
[34] and recent GPS extension to hollow resonance series in Li [49], a detailed and 
elaborate study has been made for the 7 positive ions in the Li-isoelectronic sequence 
(Z=4--10). The nonrelativistic excitation energies and radial densities were presented for 
both low and higher members in these odd- and even-parity Rydberg series with excellent 
accuracy. First all the 8 n=2 intrashell triply excited states, {\em viz.,} 2s$^2$2p 
$^2$P$^o$; 2s2p$^2$ $^2$D$^e$, $^4$P$^e$, $^2$P$^e$, $^2$S$^e$; and 2p$^3$ $^2$D$^o$, 
$^2$P$^o$, $^4$S$^o$; of B$^{2+}$, N$^{4+}$ and F$^{6+}$ were compared with the available 
literature data. Then we compared the calculated excitation energies of 8 $2l2l'$n$l''$ 
(3$\leq$n$\leq$6) resonance series of all the ions with the only reference TDM result 
available so far in the literature. Finally we singled out the 2s$^2$ns $^2$S$^e$ series to
demonstrate the feasibility and viability of this formalism for the high-lying Rydberg series
by calculating the latter for all members with n up to a maximum of 25. Literature results 
for these systems have been recently reported for a maximum n of 17 for F$^{6+}$ ($\leq 16$
for other ions) and thus all the resonances lying above are presented here for the first 
time. Out of about 380 states considered in this work, the excitation energies show a maximum
absolute deviation of 0.57\% (for 2s$^2$3p $^2$P$^o$ state of Be$^+$; excepting this lone 
case the latter remains below 0.50\%), which further consolidates the validity of the present 
density functional formalism in the domain of atomic excited states. Although a substantial
amount of results are currently available for the higher resonances in Li, the same for 
the higher members of the isoelectronic sequence are very rare, and we hope that these 
results will provide helpful references in future theoretical and experimental works. 
Naturally more accurate theoretical calculations and experimental measurements would be 
required which could shed more light on our understanding of these triply excited resonances.

The use of local nonvariational work-function based exchange potential and the nonlocal 
LYP correlation energy functional helps retaining the features of traditional wave function
based quantum mechanics within DFT, that lies at the success of this approach. The 
nonuniform spatial GPS discretization of the pertinent KS-type equation in conjunction 
with the nonlinear mapping and the symmetrization procedure within an essentially single 
determinantal framework, offers results that are quite comparable in accuracy to other 
sophisticated and elaborate theoretical methodologies existing in the literature. Given the 
fact that these are strongly correlated systems, and the work-function exchange potential
offers almost HF-quality results for ground and excited states of a variety of physical 
systems [42,72,73] including atoms, ions, metals etc., a major and obvious source of error
in our work is the correlation energy functional, which is designed mainly for the ground
states. While we are forced by the current inadequate status of DFT to use this potential,
development of a universal prescription for the correlation potential, if possible, in the 
same spirit as the work-function exchange, would be highly desirable and challenging. Put 
differently, the correlation functional may be either improved or replaced to describe the 
intricate correlation effects in a transparent way leading to better accuracy. A secondary 
and lesser contributing source of inaccuracy may be due to the tacit assumption of spherical
symmetry in the work-function exchange. The extension of this method to other hollow 
resonances, as well to higher photon-energy hollow series where all the three electrons 
reside in the shells with $n\ge4$ are equally straightforward and simple. In conclusion, 
this work emphasizes the applicability and feasibility of a simple general and efficient 
DFT-based formalism for the accurate and reliable calculation of the hollow resonances in 
multiply excited atomic systems.

\begin{acknowledgments}
I gratefully acknowledge the warm hospitality provided by the University of New 
Brunswick, Fredericton, Canada. 
\end{acknowledgments}


\begin{thebibliography}{99}
\bibitem{1}C.\ E.\ Kuyatt, J.\ A.\ Simpson and S.\ R.\ Mielczarek, Phys.\ Rev.\ A
\textbf{138}, A385 (1965).
\bibitem{2}R.\ Bruch, G.\ Paul, J.\ Andr\"{a} and L.\ Lipsky, Phys.\ Rev.\ 
A \textbf{12}, 1808 (1975).
\bibitem{3}M.\ Rodbro, R.\ Bruch and P.\ Bisgaard, J.\ Phys.\ B \textbf{12},
2413 (1979).
\bibitem{4}L.\ M.\ Kiernan, E.\ T.\ Kennedy, J.-P.\ Mosnier, J.\ T.\ Costello and 
B.\ F.\ Sonntag, Phys.\ Rev.\ Lett.\ \textbf{72}, 2359 (1994).
\bibitem{5}L.\ M.\ Kiernan, M.-K.\ Lee, B.\ F.\ Sonntag, P.\ Sladeczek, P.\ 
Zimmermann, E.\ T.\ Kennedy, J.-P.\ Mosnier and J.\ T.\ Costello, J.\ Phys.\ B
\textbf{28}, L161 (1995). 
\bibitem{6}Y.\ Azuma, F.\ Koike, J.\ W.\ Cooper, T.\ Nagata, G.\ Kutluk, E.\ 
Shigemasa, R.\ Wehlitz and I.\ A.\ Sellin, Phys.\ Rev.\ Lett.\ \textbf{79}, 
2419 (1997).
\bibitem{7}D.\ Cubaynes, S.\ Diehl, L.\ Journel, B.\ Rouvellou, J.-M.\ Bizau,
S.\ Al Moussalami, F.\ J.\ Wuilleumier, N.\ Berrah, L.\ Vo Ky, P.\ Faucher,
A.\ Hibbert, C.\ Blancard, E.\ T.\ Kennedy, T.\ J.\ Morgan, J.\ Bozek and
A.\ S.\ Schlachter, Phys.\ Rev.\ Lett.\ \textbf{77}, 2194 (1996).
\bibitem{8}S.\ Diehl, D.\ Cubaynes, F.\ J.\ Wuilleumier, J.-M.\ Bizau, L.\ 
Journel, E.\ T.\ Kennedy, C.\ Blancard, L.\ Vo Ky, P.\ Faucher, A.\ Hibbert, N.\ 
Berrah, T.\ J.\ Morgan, J.\ Bozek and A.\ S.\ Schlachter, Phys.\ Rev.\ Lett.\ 
\textbf{79}, 1241 (1997).
\bibitem{9}S.\ Diehl, D.\ Cubaynes, H.\ L.\ Zhou, L.\ Vo Ky, F.\ J.\ Wuilleumier,
E.\ T.\ Kennedy, J.-M.\ Bizau, S.\ T.\ Manson, T.\ J.\ Morgan, C.\ Blancard, N.\
Berrah and J.\ Bozek, Phys.\ Rev.\ Lett.\ \textbf{84}, 1677 (2000).
\bibitem{10}L.\ Journel, D.\ Cubaynes, J.-M.\ Bizau, S.\ Al Moussalami, B.\ 
Rouvellou, F.\ J.\ Wuilleumier, L.\ Vo Ky, P.\ Faucher and A.\ Hibbert, Phys.\ 
Rev.\ Lett.\ \textbf{76}, 30 (1996).
\bibitem{11}S.\ Diehl, D.\ Cubaynes, E.\ T.\ Kennedy, F.\ J.\ Wuilleumier,
J.-M.\ Bizau, L.\ Journel, L.\ Vo Ky, P.\ Faucher, A.\ Hibbert, C.\ Blancard,
N.\ Berrah, T.\ J.\ Morgan, J.\ Bozek and A.\ S.\ Schlachter, J.\ Phys.\ B
\textbf{30}, L595 (1997).
\bibitem{12}S.\ Diehl, D.\ Cubaynes, K.\ T.\ Chung, F.\ J.\ Wuilleumier, E.\ T.\ 
Kennedy, J.-M.\ Bizau, L.\ Journel, C.\ Blancard, L.\ Vo Ky, P.\ Faucher,
A.\ Hibbert, N.\ Berrah, T.\ J.\ Morgan, J.\ Bozek and A.\ S.\ Schlachter, Phys.\
Rev.\ A \textbf{56}, R1071 (1997).
\bibitem{13}M.\ Ahmed and L.\ Lipsky, Phys.\ Rev.\ A \textbf{12}, 1176 (1975).
\bibitem{14}U.\ I.\ Safronova and V.\ S.\ Senashenko, J.\ Phys.\ B \textbf{11},
2623 (1978).
\bibitem{15}K.\ T.\ Chung and B.\ C.\ Gou, Phys.\ Rev.\ A \textbf{52}, 3669 (1995).
\bibitem{16}K.\ T.\ Chung and B.\ C.\ Gou, Phys.\ Rev.\ A \textbf{53}, 2189 (1996).
\bibitem{17}B.\ C.\ Gou and K.\ T.\ Chung, J.\ Phys.\ B \textbf{29}, 6103 (1996).
\bibitem{18}B.\ C.\ Gou, Eur.\ Phys.\ J.\ D \textbf{5}, 39 (1999). 
\bibitem{19}N.\ A.\ Piangos and C.\ A.\ Nicolaides, Phys.\ Rev.\ A \textbf{48},
4142 (1993).
\bibitem{20}C. A.\ Nicolaides and N.\ A.\ Piangos, J.\ Phys.\ B \textbf{34}, 99 
(2001).
\bibitem{21}H.\ Bachau, J.\ Phys.\ B \textbf{29}, 4365 (1996).
\bibitem{22}K.\ Berrington and S.\ Nakazaki, J.\ Phys.\ B \textbf{31}, 313 (1998).
\bibitem{23}H.\ L.\ Zhou, S.\ T.\ Manson, P.\ Faucher and L.\ Vo Ky, Phys.\ Rev.\ A
\textbf{62}, 012707 (2000).
\bibitem{24}M.\ J.\ Conneely and L.\ Lipsky, Phys.\ Rev.\ A \textbf{61}, 032506 
(2000).
\bibitem{25}M.\ J.\ Conneely and L.\ Lipsky, At.\ Data Nucl.\ Data Tables
\textbf{82}, 115 (2002).
\bibitem{26}M.\ J.\ Conneely and L.\ Lipsky, At.\ Data Nucl.\ Data Tables
\textbf{86}, 35 (2004).
\bibitem{27}T.\ Morishita and C.\ D.\ Lin, Phys.\ Rev.\ A \textbf{59}, 1835 (1999).
\bibitem{28}T.\ Morishita and C.\ D.\ Lin, Phys.\ Rev.\ A \textbf{67}, 022511 (2003).
\bibitem{29}U.\ I.\ Safronova and R.\ Bruch, Phys.\ Scr.\ \textbf{57}, 519 (1998).
\bibitem{30}N.\ Vacek and J.\ E.\ Hansen, J.\ Phys.\ B \textbf{25}, 883 (1992).
\bibitem{31}L.\ B.\ Madsen and K.\ M\o lmer, Phys.\ Rev.\ A \textbf{65}, 022506 (2002).
\bibitem{32}L.\ B.\ Madsen and K.\ M\o lmer, J.\ Phys.\ B \textbf{36}, 769 (2003).
\bibitem{33} N.\ A.\ Piangos and C.\ A.\ Nicolaides, Phys.\ Rev.\ A \textbf{67}, 
052501 (2003).
\bibitem{34}A.\ K.\ Roy, R.\ Singh and B.\ M.\ Deb, Int.\ J.\ Quant.\ Chem.\ 
\textbf{65}, 317 (1997).
\bibitem{35}M.\ K.\ Harbola and V.\ Sahni, Phys.\ Rev.\ Lett. \textbf{62}, 
489 (1989).
\bibitem{36}C.\ Lee, W.\ Yang and R.\ G.\ Parr, Phys.\ Rev.\ B \textbf{37}, 785
(1988).
\bibitem{37}R.\ Singh and B.\ M.\ Deb, J.\ Chem.\ Phys.\ \textbf{104}, 5892 (1996).
\bibitem{38}A.\ K.\ Roy, R.\ Singh and B.\ M.\ Deb, J.\ Phys.\ B \textbf{30},
4763 (1997).
\bibitem{39}A.\ K.\ Roy and B.\ M.\ Deb, Phys.\ Lett.\ A \textbf{234}, 465 (1998).
\bibitem{40}A.\ K.\ Roy and B.\ M.\ Deb, Chem.\ Phys.\ Lett.\ \textbf{292}, 
461 (1998).
\bibitem{41}R.\ Singh, A.\ K.\ Roy and B.\ M.\ Deb, Chem.\ Phys.\ Lett.\ 
\textbf{296}, 530 (1998). 
\bibitem{42}R.\ Singh and B.\ M.\ Deb, Phys.\ Rep.\ \textbf{311}, 47 (1999).
\bibitem{43}A.\ K.\ Roy and S.\ I.\ Chu, Phys.\ Rev.\ A \textbf{65}, 052508 (2002).
\bibitem{44}X.\ M.\ Tong and S.\ I.\ Chu, Phys.\ Rev.\ A \textbf{64}, 013417 (2001).
\bibitem{45}D.\ Telnov and S.\ I.\ Chu, Phys.\ Rev.\ A \textbf{66}, 043417 (2002).
\bibitem{46}A.\ K.\ Roy and S.\ I.\ Chu, Phys.\ Rev.\ A \textbf{66}, 043402 (2002).
\bibitem{47}A.\ K.\ Roy, Phys.\ Lett.\ A \textbf{321}, 231 (2004).
\bibitem{48}A.\ K.\ Roy, J.\ Phys.\ G \textbf{30}, 269 (2004).
\bibitem{49}A.\ K.\ Roy, J.\ Phys.\ B \textbf{37}, 4369 (2004).  
\bibitem{50}P.\ Hohenberg and W.\ Kohn, Phys.\ Rev.\ \textbf{136}, B864 (1964).
\bibitem{51}W.\ Kohn and L.\ J.\ Sham, Phys.\ Rev.\ \textbf{140}, A1133 (1965).
\bibitem{52}B.\ M.\ Deb and S.\ K.\ Ghosh, in {\em  The Single-Particle Density in Physics
and Chemistry}, N.\ H.\ March and B.\ M.\ Deb (eds.) (Academic Press, London, 1987).
\bibitem{53}A.\ K.\ Theophilou, J.\ Phys.\ C \textbf{12}, 5419 (1979).
\bibitem{54}N.\ Hadjisavvas and A.\ K.\ Theophilou, Phys.\ Rev.\ A \textbf{30}, 2183 
(1984); {\em ibid.} \textbf{32}, 720 (1985).
\bibitem{55}\'{A}.\ Nagy, Int.\ J.\ Quant.\ Chem.\ \textbf{69}, 247 (1998).
\bibitem{56}F.\ Tasn\'{a}di and \'{A}.\ Nagy, Int.\ J.\ Quant.\ Chem.\ \textbf{92}, 
234 (2003). 
\bibitem{57}E.\ K.\ U.\ Gross, L.\ N.\ Oliveira and W.\ Kohn, Phys.\ Rev.\ A 
\textbf{37}, 2805 (1988); {\em ibid.} \textbf{37}, 2809 (1988).
\bibitem{58}M.\ Petersilka, U.\ J.\ Gossmann and E.\ K.\ U.\ Gross, Phys.\ Rev.\ 
Lett.\ \textbf{76}, 1212 (1996).
\bibitem{59}C.\ Jamorski, M.\ E.\ Casida and D.\ R.\ Salahub, J.\ Chem.\ Phys.\ 
\textbf{104}, 5134 (1996).
\bibitem{60}E.\ San Fabi\'{a}n and L.\ Pastor-Abia, Int.\ J.\ Quant.\ Chem.\ 
\textbf{91}, 451 (2003).
\bibitem{61}A.\ G\"{o}rling, Phys.\ Rev.\ A \textbf{54}, 3912 (1996). 
\bibitem{62}C.\ Filippi, C.\ Umrigar and X.\ Gonze, J.\ Chem.\ Phys.\ \textbf{107}, 
9994 (1997).
\bibitem{63}T.\ Mineva, A.\ Goursot and C.\ Daul, Chem.\ Phys.\ Lett.\ \textbf{350}, 
147 (2001). 
\bibitem{64}J.\ C.\ Slater, {\em Quantum Theory of Atomic Structure} (McGraw Hill, New
York, 1960), Vol. II.
\bibitem{65}G.\ Yao and S.\ I.\ Chu, Chem.\ Phys.\ Lett.\ \textbf{204}, 381 (1993).
\bibitem{66}J.\ Wang, S.\ I.\ Chu and C.\ Laughlin, Phys.\ Rev.\ A \textbf{50}, 3208
(1994).
\bibitem{67}Z.\-C.\ Yan, M.\ Tambasco and G.\ W.\ F.\ Drake, Phys.\ Rev.\ A, \textbf{57},
1652 (1998).
\bibitem{68}K.\ T.\ Chung, Phys.\ Rev.\ A \textbf{44}, 5421 (1991). 
\bibitem{69}F.\ W.\ King, Phys.\ Rev.\ A \textbf{38}, 6017 (1988).
\bibitem{70}F.\ W.\ King, Phys.\ Rev.\ A \textbf{40}, 1735 (1989).
\bibitem{71}B.\ F.\ Davis and K.\ T.\ Chung, Phys.\ Rev.\ A \textbf{42}, 5121 (1990).
\bibitem{72}V.\ Sahni, in {\em Structure and Dynamics of Atoms and Molecules: Conceptual
Trends}, J.\ L.\ Calais and E.\ S.\ Kryachko, (eds.) (Kluwer Academic, Dordrecht, 1995).
\bibitem{73}V.\ Sahni and A.\ Solomatin, Adv.\ Quant.\ Chem.\ \textbf{33}, 241 (1999).
\end{thebibliography}
\end{document}